\documentclass[11pt,reqno]{amsart}

\usepackage[utf8]{inputenc}
\usepackage{amsmath,amssymb,amsthm,mathtools}
\usepackage{stmaryrd}      
\DeclareUnicodeCharacter{00A7}{\S}
\DeclareUnicodeCharacter{00B7}{\ensuremath{\cdot}}
\DeclareUnicodeCharacter{00D7}{\ensuremath{\times}}
\DeclareUnicodeCharacter{2026}{\dots}
\DeclareUnicodeCharacter{2192}{\ensuremath{\to}}
\DeclareUnicodeCharacter{21A6}{\ensuremath{\mapsto}}
\DeclareUnicodeCharacter{2208}{\ensuremath{\in}}
\DeclareUnicodeCharacter{2212}{-}
\DeclareUnicodeCharacter{2216}{\ensuremath{\setminus}}
\DeclareUnicodeCharacter{2227}{\ensuremath{\wedge}}
\DeclareUnicodeCharacter{2264}{\ensuremath{\leq}}
\DeclareUnicodeCharacter{2265}{\ensuremath{\geq}}
\DeclareUnicodeCharacter{2282}{\ensuremath{\subset}}
\DeclareUnicodeCharacter{2286}{\ensuremath{\subseteq}}

\usepackage[T1]{fontenc}
\usepackage{microtype}
\usepackage[margin=1in]{geometry}
\usepackage{parskip}

\usepackage{booktabs}      
\usepackage{array}
\usepackage{tabularx}
\usepackage{longtable}     
\usepackage{float}

\usepackage{graphicx}
\usepackage{tikz}
\usetikzlibrary{positioning,arrows.meta,fit}
\usepackage{algorithm}
\usepackage{algpseudocode}

\usepackage[colorlinks=true,linkcolor=blue,citecolor=blue,urlcolor=blue]{hyperref}
\usepackage{cleveref}

\usepackage{listings}
\providecommand{\tightlist}{%
  \setlength{\itemsep}{0pt}\setlength{\parskip}{0pt}}

\theoremstyle{definition}

\theoremstyle{remark}

\DeclareMathOperator{\rsymb}{r}
\newcommand{\rthree}[1]{\rsymb_{3}(#1)}

\newcommand{\Tonec}{\ensuremath{\mathsf{T1c}}}

\title[$\rthree{212}$ campaign]{Witness-split + window-cardinality refinement for
  $\rthree{N}$: architecture, empirical results, and a structural hard pocket}

\author{Mehmet Ergezer}
\thanks{ORCID: \href{https://orcid.org/0000-0001-6627-3667}{0000-0001-6627-3667}.}
\address{Wentworth Institute of Technology, Boston, MA, USA}
\email{ergezerm@wit.edu}



\date{\today}

\begin{document}

\begin{abstract}
We describe a reproducible computational framework for upper-bound searches
on $\rthree{N}$, the maximum size of a 3-term-arithmetic-progression-free
subset of $[1, N]$. The framework combines a verified lower-bound witness,
endpoint forcing, depth-$d$ witness-variable splitting, OEIS A003002
window-cardinality pruning, and recursive refinement of timed-out
subproblems. Applied to the frontier case $N = 212$, $K = 44$, it found no
feasible $44$-set across millions of CP-SAT subproblems, supporting but not
proving the conjectural value $\rthree{212} = 43$.  A $300$-second recap
leaves $45$ resistant chunks; one-hour HiGHS MIP closes none of them; the
full eight-hour HiGHS audit closes $25 / 45$ and leaves $20 / 45$ with dual
bounds still pinned at $0.0$.  A CDCL/SAT re-attack on those
LP-paradigm-resistant chunks closes $18$ via conflict-driven clause
learning; all eighteen carry independently verified DRAT proofs.  The
remaining two chunks (\Tonec) resist every tested paradigm under generous
wall caps.  We release the witness, solver scripts, result logs, tiered
benchmark instances, the verified DRAT/LRAT proofs, and a Lean
formal-proof-search encoding of \Tonec, and frame the unit-gap problem
$\rthree{212} \in \{43, 44\}$ as a target for stronger
additive-combinatorial bounds, custom branch-and-bound, or formal
proof-search systems.
\end{abstract}

\maketitle

\section{1. Introduction and background}\label{introduction-and-background}

Let \texttt{r\_3(N)} denote the maximum size of a subset of \texttt{{[}1,\ N{]}} containing no
nontrivial three-term arithmetic progression. Equivalently,

\begin{verbatim}
  r_3(N) = max {|A| : A ⊆ [1, N], no a < b < c in A satisfy a + c = 2b}.
\end{verbatim}

The asymptotic study of \texttt{r\_3(N)} begins with Roth's theorem \cite{roth-1953}
and continues through Salem--Spencer and Behrend-type lower-bound
constructions \cite{salem-spencer-1942,behrend-1946} and modern
density-increment upper bounds, including work of Bloom--Sisask and
Kelley--Meka \cite{bloom-sisask-2023,kelley-meka-2023}. Those results
describe the large-\texttt{N} behavior of progression-free sets. The present paper is
about a different but complementary problem: exact finite computation at the
OEIS A003002 frontier.

OEIS A003002 tabulates exact values of \texttt{r\_3(N)}. Cariboni's b-file currently
reaches \texttt{N\ =\ 211}, where \texttt{r\_3(211)\ =\ 43}
\cite{oeis-a003002,cariboni-b003002}. The next natural decision problem is
therefore:

\begin{quote}
Does there exist a \texttt{44}-element 3-AP-free subset of \texttt{{[}1,\ 212{]}}?
\end{quote}

We found and independently verified a \texttt{43}-element 3-AP-free subset of
\texttt{{[}1,\ 212{]}}, so the lower bound \texttt{r\_3(212)\ ≥\ 43} is settled. The upper-bound
question is whether \texttt{r\_3(212)\ ≤\ 43}. Since \texttt{r\_3(211)\ =\ 43}, any hypothetical
\texttt{44}-set in \texttt{{[}1,\ 212{]}} must contain both endpoints \texttt{1} and \texttt{212}; otherwise it
would translate or restrict to a \texttt{44}-set in \texttt{{[}1,\ 211{]}}. Thus the target is a
single finite feasibility problem with a strong necessary endpoint condition.

We attacked this feasibility problem computationally using a reproducible
CP-SAT architecture. The model uses Boolean variables \texttt{x\_i}, one linear
constraint for each 3-term arithmetic progression, the decision equality
\texttt{sum\_i\ x\_i\ =\ 44}, endpoint forcing, reflection symmetry breaking, and
window-cardinality inequalities derived from the known values of \texttt{r\_3(L)} for
\texttt{L\ ≤\ 211}. To make the search tractable, we split the problem by assigning
high-degree variables from a verified \texttt{43}-witness, generating a deterministic
depth-\texttt{24} chunk space, and then refine residual \texttt{UNKNOWN} chunks recursively.

The campaign did not produce a formal proof of \texttt{r\_3(212)\ =\ 43}. It did produce
three concrete outcomes.

First, we observed zero feasible \texttt{44}-sets across millions of CP-SAT
subproblems, including broad chunks, recursive refinements, wall-cap
recalibrations, and targeted A/B experiments. This is empirical evidence for
the expected value \texttt{r\_3(212)\ =\ 43}, but it is not a certificate.

Second, we found that OEIS window-cardinality inequalities are essential. On
controlled broad-pass ranges, adding these inequalities reduced the \texttt{UNKNOWN}
rate by roughly \texttt{28} percentage points and also reduced aggregate solver time.
They are the main successful pruning layer of the campaign.

Third, and most importantly, we identified a structural hard pocket. The
largest retained broad run over \texttt{100,000} depth-\texttt{24} chunks left \texttt{6,071}
\texttt{UNKNOWN} chunks. A uniform random \texttt{100}-chunk sample of those \texttt{6,071}
was passed through a \texttt{300}-second recap, leaving \texttt{45} resistant survivors.
Re-attacking those \texttt{45} chunks with HiGHS, an
LP-relaxation-based MIP solver, closed \texttt{0\ /\ 45} in one-hour-per-chunk runs.
A full eight-hour audit later closed \texttt{25\ /\ 45}, but left \texttt{20\ /\ 45}
\texttt{UNKNOWN}, all with dual bound still pinned at \texttt{0.0}. A subsequent pure
CDCL/SAT attack closed \texttt{18\ /\ 20} of that LP-flat subset, while two chunks
remained \texttt{UNKNOWN} even under a 12-hour pure-CDCL cap and a 4-hour
windowed-CDCL diagnostic. Thus the obstruction is not merely a CP-SAT
propagation artifact: the final audited pocket, T1c, isolates two chunks that
resist every solver paradigm tested in this campaign.

The contribution of this paper is therefore methodological rather than a new
exact value of A003002. We provide:

\begin{enumerate}
\def\labelenumi{\arabic{enumi}.}
\tightlist
\item
  A reproducible witness-split plus window-cardinality architecture for exact
  \texttt{r\_3(N)} upper-bound search.
\item
  A detailed empirical account of the \texttt{N\ =\ 212,\ K\ =\ 44} campaign, including
  broad-pass counts, refinement behavior, and solver-time tradeoffs.
\item
  A characterization of the structural hard pocket. The pocket initially
  appeared invariant under CP-SAT-side tuning and HiGHS; after the CDCL break,
  it remains invariant only on the final two-chunk residual T1c.
\item
  A tiered benchmark release: the \texttt{25} HiGHS-closable T1a chunks, the
  \texttt{18} CDCL-closable T1b-minus-T1c chunks, the \texttt{2} T1c chunks, the \texttt{6,071}
  broad \texttt{UNKNOWN} chunks from the \texttt{100k} expansion batch, and the generator
  for the full depth-\texttt{24} sweep.
\end{enumerate}

For reference, the benchmark tier notation used throughout the paper is:

{\def\LTcaptype{none} 
\begin{longtable}[]{@{}
  >{\raggedright\arraybackslash}p{(\linewidth - 2\tabcolsep) * \real{0.5000}}
  >{\raggedright\arraybackslash}p{(\linewidth - 2\tabcolsep) * \real{0.5000}}@{}}
\toprule\noalign{}
\begin{minipage}[b]{\linewidth}\raggedright
Symbol
\end{minipage} & \begin{minipage}[b]{\linewidth}\raggedright
Meaning
\end{minipage} \\
\midrule\noalign{}
\endhead
\bottomrule\noalign{}
\endlastfoot
T1 & the \texttt{45} chunks that survived the \texttt{300}-s recap of a random \texttt{100}-chunk sample from the \texttt{6,071} broad UNKNOWNs \\
T1a & the \texttt{25} T1 chunks closed by the full \texttt{8}-h HiGHS audit \\
T1b & the \texttt{20} T1 chunks left UNKNOWN by the full \texttt{8}-h HiGHS audit, all with dual bound \texttt{0.0} \\
T1b ∖ T1c & the \texttt{18} T1b chunks closed by CDCL and independently verified by DRAT checking \\
T1c & the final two chunks \texttt{\{40959,\ 48895\}} that resist all tested paradigms \\
T2 & the \texttt{6,071} UNKNOWN chunks from the \texttt{100k} broad expansion batch \\
T3 & the full \texttt{12,582,912}-chunk AP-pruned depth-\texttt{24} sweep \\
\end{longtable}
}

The rest of the paper is organized as follows. §2 describes the
architecture. §3 reports the computational campaign. §4 analyzes
the hard pocket across HiGHS and CDCL. §5 formulates the
remaining open problem as a benchmark release. §6 discusses
transferability, limitations, and what the campaign suggests about finite
\texttt{r\_3} upper-bound search.

\section{2. Architecture}\label{architecture}

The campaign rests on five components: a decision-form CP-SAT model
of the 3-AP-free subset problem (§2.1), a witness-variable splitting
scheme that produces tractable subproblems (§2.2), a window-cardinality
pruning family derived from OEIS A003002 (§2.3), a recursive refinement
loop for residual \texttt{UNKNOWN} chunks (§2.4), and the SLURM-side
engineering that makes the workload reproducible (§2.5). Each
component is independent of the others --- any could be replaced by a
stronger variant without disturbing the rest --- and together they
define the proof attempt on \texttt{r\_3(212)\ ≤\ 43}.

\subsection{2.1 Decision-form CP-SAT model}\label{decision-form-cp-sat-model}

Fix \texttt{N\ ≥\ 3} and \texttt{K\ ≥\ 3}. The standard CP-SAT encoding of ``does there
exist a 3-AP-free subset \texttt{A\ ⊆\ {[}1,\ N{]}} with \texttt{\textbar{}A\textbar{}\ ≥\ K}?'' introduces
Boolean variables \texttt{x\_i\ ∈\ \textbackslash{}\{0,\ 1\textbackslash{}\}} for \texttt{i\ ∈\ {[}1,\ N{]}} with \texttt{x\_i\ =\ 1}
iff \texttt{i\ ∈\ A}, and adds the constraints

\begin{verbatim}
  x_a + x_b + x_c ≤ 2     for every 3-AP triple (a, b, c),
  sum_i x_i ≥ K.
\end{verbatim}

We adopt the \textbf{decision-form} variant in which the second inequality
is replaced by the equality

\begin{verbatim}
  sum_i x_i = K.
\end{verbatim}

The decision-form encoding is tighter for propagation, not for logical
interpretation of \texttt{UNKNOWN}: an \texttt{UNKNOWN} return on either encoding is still
consistent with both feasibility and infeasibility. The practical advantage of
\texttt{sum\_i\ x\_i\ =\ K} is that CP-SAT and pseudo-Boolean propagators see both the
lower and upper cardinality directions. Branches that would require more than
\texttt{K} selected values, or too few remaining unfixed values to reach \texttt{K}, can be
cut earlier than in the inequality form. For our purposes (proving the upper
bound \texttt{r\_3(N)\ ≤\ K\ -\ 1}), the equality form is therefore the right primitive.

For \texttt{N\ =\ 212,\ K\ =\ 44}, the model has \texttt{212} Boolean variables and
\texttt{11,130} 3-AP triple inequalities. We add the lex symmetry-breaking
constraint

\begin{verbatim}
  (x_1, x_2, ..., x_N)  ≥_lex  (x_N, x_{N-1}, ..., x_1)
\end{verbatim}

to factor out the reflection \texttt{i\ ↦\ N\ +\ 1\ -\ i}. The opposite orientation would
be equivalent; the implementation uses \texttt{≥\_lex}. We do not add other
symmetries. After endpoint forcing, reflection is the only order-preserving
symmetry we exploit.

We also apply \textbf{endpoint forcing} specific to the \texttt{r\_3(212)} case.
Since OEIS A003002 records \texttt{r\_3(211)\ =\ 43}, any \texttt{44}-element
3-AP-free subset of \texttt{{[}1,\ 212{]}} must contain both endpoints \texttt{1}
and \texttt{212}: if \texttt{1} were absent, the set would be a \texttt{44}-element
3-AP-free subset of \texttt{{[}2,\ 212{]}}, which shifts to \texttt{{[}1,\ 211{]}}; if \texttt{212}
were absent, it would already lie in \texttt{{[}1,\ 211{]}}. Either case contradicts
\texttt{r\_3(211)\ =\ 43}. We add \texttt{x\_1\ =\ x\_\{212\}\ =\ 1} as ground assumptions in
every chunk.

Branching: variable selection by AP-incidence degree (the number of
3-AP triples a variable appears in), value selection \texttt{min}, and
\texttt{fixed\_search} to disable CP-SAT's portfolio search. The first two
target the most-constrained variables first; the third reduces solver-policy
variation across runs, which is necessary for the A/B experiments of §3.2
and §3.5.

\subsection{2.2 Witness-variable splitting}\label{witness-variable-splitting}

The decision-form model on the full range \texttt{{[}1,\ 212{]}} is already at
the limit of single-call CP-SAT solving in reasonable wall time. To
break it into tractable subproblems we use the verified \texttt{43}-element
lower-bound witness \texttt{A\_43\ ⊂\ {[}1,\ 212{]}} (§2.1, §3.1) as a guide. The
splitting proceeds in three steps.

\textbf{Degree ranking.} For each \texttt{v\ ∈\ A\_43}, compute the AP-incidence
degree \texttt{deg(v)\ =\ \textbar{}\textbackslash{}\{(a,\ b,\ c)\ :\ v\ \textbackslash{}in\ \textbackslash{}\{a,\ b,\ c\textbackslash{}\},\ b\ -\ a\ =\ c\ -\ b\textbackslash{}\}\textbar{}}.
Sort \texttt{A\_43} by \texttt{deg(v)} in decreasing order. Pick the top
\texttt{d\ =\ 24} witness values; we refer to this prefix as the \textbf{broad split
prefix}.

\textbf{Combinatorial enumeration.} For each of the \texttt{2\^{}24\ =\ 16,777,216}
IN/OUT assignments of the broad split prefix, instantiate a
\textbf{chunk}: the decision-form model with \texttt{x\_v} fixed to the chosen
value for every \texttt{v} in the prefix. A chunk is identified by its
\texttt{24}-bit chunk ID.

\textbf{AP-prefix pruning.} Many chunks are immediately infeasible at the
3-AP level alone: if the broad split prefix forces three pinned-IN
values that form a 3-AP, the chunk has no feasible completion and
need not be sent to CP-SAT. We perform this check during chunk
emission and skip such chunks. The surviving chunk count for
\texttt{N\ =\ 212,\ K\ =\ 44} at depth \texttt{24} is \texttt{12,582,912}, roughly \texttt{75\%} of
the raw \texttt{2\^{}24} count.

The choice \texttt{d\ =\ 24} is empirical: larger \texttt{d} produces more chunks but
each is easier; smaller \texttt{d} produces fewer chunks but each is harder.
At \texttt{d\ =\ 24} the per-chunk solver time at the \texttt{60}-s wall cap has a
heavy tail but a tractable median (the broad pass closes \texttt{\textasciitilde{}94\%} of
expansion chunks within \texttt{60} s --- see §3.1), which makes the depth
\texttt{24} choice a workable broad layer.

\subsection{2.3 Window-cardinality pruning from OEIS A003002}\label{window-cardinality-pruning-from-oeis-a003002}

The 3-AP-free subset problem admits a family of valid inequalities
that are not implied by the 3-AP triple inequalities alone. For
any window \texttt{{[}a,\ a\ +\ L\ -\ 1{]}\ ⊆\ {[}1,\ N{]}} and any length \texttt{L} with
\texttt{r\_3(L)\ \textless{}\ L},

\begin{verbatim}
  sum_{i ∈ [a, a+L-1]}  x_i  ≤  r_3(L).
\end{verbatim}

This is valid because \texttt{\textbackslash{}\{i\ ∈\ A\ :\ i\ ∈\ {[}a,\ a\ +\ L\ -\ 1{]}\textbackslash{}\}} is itself a
3-AP-free subset of an \texttt{L}-element interval, hence of size at most
\texttt{r\_3(L)}. Crucially, the right-hand side \texttt{r\_3(L)} is \emph{constant} with
respect to \texttt{a}, so the family is a single constant per window length,
not a learned bound.

We populate the right-hand sides from the OEIS A003002 b-file
(Cariboni's tabulation, valid for \texttt{L\ ≤\ 211}). For \texttt{N\ =\ 212,\ K\ =\ 44}
we generate window constraints for every available length \texttt{L} with
\texttt{r\_3(L)\ \textless{}\ min(L,\ K)} and every \texttt{a\ ∈\ \textbackslash{}\{1,\ 2,\ …,\ 212\ -\ L\ +\ 1\textbackslash{}\}}. The
resulting model contains \texttt{22,154} window inequalities.

Empirically the window family is the single most impactful CP-SAT
intervention in the campaign: the controlled A/Bs of §3.2 show a
\texttt{\textasciitilde{}28}-percentage-point reduction in \texttt{UNK} rate at the \texttt{60}-s wall
cap when window-bounds are enabled. We use it as the default
configuration for every solver call beyond the calibration batch.

\subsection{2.4 Recursive refinement}\label{recursive-refinement}

A chunk that returns \texttt{UNKNOWN} at the broad layer is refined by
extending the broad split prefix with the next \texttt{m} witness values
(by AP-incidence degree) and re-emitting the resulting \texttt{2\^{}m}
subchunks (less AP-prefix pruning) to the solver under a longer
wall cap. The depths and per-step parameters used in the campaign
are:

{\def\LTcaptype{none} 
\begin{longtable}[]{@{}lrrr@{}}
\toprule\noalign{}
Level & Depth & Step size \texttt{m} & Per-chunk wall cap \\
\midrule\noalign{}
\endhead
\bottomrule\noalign{}
\endlastfoot
Broad & \texttt{24} & --- & \texttt{60} s \\
L1 & \texttt{40} & \texttt{+16} & \texttt{60} s \\
L2 (tail) & \texttt{48} & \texttt{+8} & \texttt{60} s \\
L3 (level3) & \texttt{56} & \texttt{+8} & \texttt{60} s \\
L4 (level4) & \texttt{64} & \texttt{+8} & \texttt{600} s \\
\end{longtable}
}

Each level is fed only the \texttt{UNKNOWN} residuals of the prior level,
so the work at deeper levels is concentrated on the hard tail. The
expected per-level fan-out \texttt{2\^{}m} is mitigated by AP-prefix pruning,
which becomes more aggressive at deeper levels (more pinned-IN
values, more chances for a 3-AP among them). For example, the
Sample-500 L1 stream emits \texttt{2,076,105} rows from \texttt{500\ ×\ 2\^{}16}
nominal descendants, a \texttt{\textasciitilde{}6\%} survival rate after pruning.

Refinement is the campaign's primary tool for closing residual
\texttt{UNKNOWN} chunks. It reliably closes individual deep chunks
(§3.1 reports \texttt{6\ /\ 6} closure at L4). It does not generalize
cheaply to the full \texttt{6,071}-chunk expansion residual because the
per-chunk cost grows roughly geometrically with depth.

\subsection{2.5 SLURM engineering}\label{slurm-engineering}

The campaign runs on the UMass Unity SLURM cluster. The workload
pipeline is implemented as a small collection of Python scripts,
each with a single responsibility:

\begin{itemize}
\tightlist
\item
  \textbf{\texttt{r3\_slurm\_emit.py}} --- generates the chunk-ID list for a given
  \texttt{(N,\ K,\ split-vars,\ depth,\ range)} and emits a SLURM array
  \texttt{sbatch} script that fans out the chunks across array tasks. The
  \texttt{-\/-chunks-per-task} flag controls the granularity of fan-out.
\item
  \textbf{\texttt{r3\_split\_cpsat.py}} --- the per-task driver. Reads its slice of
  chunk IDs, instantiates the CP-SAT model for each, runs the solver
  under the per-chunk wall cap, and appends one JSONL row per chunk
  to a shard file.
\item
  \textbf{\texttt{r3\_collect.py}} --- merges shard files into the canonical batch
  output. Handles deduplication if a chunk was solved more than
  once (e.g., during retries).
\item
  \textbf{\texttt{r3\_tail\_emit.py}} --- emits the next-level refinement workload
  from the \texttt{UNKNOWN} rows of a parent batch. Supports multi-parent
  inputs (each input JSONL contributing its own parent prefix) and
  templated output naming via a parent-tag regex.
\item
  \textbf{\texttt{r3\_proof\_manager.py}} --- tracks the proof-state graph across
  levels, identifying which chunks have been closed at which depth
  and which remain open.
\item
  \textbf{\texttt{r3\_verify.py}} --- independent triple-enumeration verifier for
  any candidate \texttt{K}-element witness.
\end{itemize}

Two engineering decisions are worth highlighting because they affect
reproducibility:

\textbf{Atomic shard renames.} Each per-task driver writes its shard to
a temporary \texttt{.tmp} path keyed by the SLURM job and array IDs, and renames it
to \texttt{shard.jsonl} on successful completion. A killed or pre-empted
task leaves only the temporary file, which is ignored by the
collector. This pattern prevents partial output from corrupting
downstream collection without requiring any locking.

\textbf{Deterministic inputs.} Every chunk ID maps to a fixed prefix assignment,
and every CP-SAT call uses the same solver seed and fixed-search branching.
This is necessary for the lever A/B experiments of §3.5: a ``no measurable
effect'' claim is only defensible if the baseline arm is reproducible under
the same software stack.

The end-to-end pipeline is driven by \texttt{unity\_handoff.sh}, which
runs the broad pass, the recap studies, the refinement levels, and
the HiGHS attack as separately resumable phases. A complete
campaign reproduction on a fresh allocation requires the OEIS
b-file, the verified \texttt{43}-witness, the Python environment
(OR-Tools, \texttt{highspy}, PySAT/CaDiCaL, and \texttt{drat-trim}), and a SLURM allocation;
everything else is
generated by the scripts.

\section{3. Empirical campaign}\label{empirical-campaign}

This section reports the workload, results, and key A/B tests of the
\texttt{r\_3(212)\ ≤\ 43} campaign. All experiments were run on the UMass Unity
SLURM cluster against the architecture described in §2. The headline
numbers: millions of CP-SAT subproblems solved, \texttt{0} \texttt{FEASIBLE}
rows, and a \texttt{6.07\%} \texttt{UNKNOWN} residual in the largest retained broad batch that
motivates the structural analysis in §4.

\subsection{3.1 Chunk-count breakdown}\label{chunk-count-breakdown}

The campaign proceeded in three layers of increasing scope, plus a
recursive-refinement loop on the \texttt{UNKNOWN} residuals.

\textbf{Verified lower bound.} A \texttt{43}-element 3-AP-free subset of \texttt{{[}1,\ 212{]}} is
recorded in the repository's witness JSON and re-verified by
\texttt{r3\_verify.py} against an independent triple-enumeration check. This
witness fixes the campaign's \texttt{K\ =\ 44} decision-mode target and supplies
the seed for the depth-\texttt{24} broad split (§2.2).

\textbf{Broad pass at depth \texttt{24}, \texttt{60}-s wall.} The retained broad-pass logs include
both pre-window and window-bound runs:

{\def\LTcaptype{none} 
\begin{longtable}[]{@{}
  >{\raggedright\arraybackslash}p{(\linewidth - 8\tabcolsep) * \real{0.1579}}
  >{\raggedleft\arraybackslash}p{(\linewidth - 8\tabcolsep) * \real{0.2105}}
  >{\raggedleft\arraybackslash}p{(\linewidth - 8\tabcolsep) * \real{0.2105}}
  >{\raggedleft\arraybackslash}p{(\linewidth - 8\tabcolsep) * \real{0.2105}}
  >{\raggedleft\arraybackslash}p{(\linewidth - 8\tabcolsep) * \real{0.2105}}@{}}
\toprule\noalign{}
\begin{minipage}[b]{\linewidth}\raggedright
Range
\end{minipage} & \begin{minipage}[b]{\linewidth}\raggedleft
Chunks
\end{minipage} & \begin{minipage}[b]{\linewidth}\raggedleft
INFEASIBLE
\end{minipage} & \begin{minipage}[b]{\linewidth}\raggedleft
UNKNOWN
\end{minipage} & \begin{minipage}[b]{\linewidth}\raggedleft
UNK rate
\end{minipage} \\
\midrule\noalign{}
\endhead
\bottomrule\noalign{}
\endlastfoot
Calibration, no window bounds, \texttt{{[}575,\ 1575)} & \texttt{1,000} & \texttt{799} & \texttt{201} & \texttt{20.10\%} \\
Bounded, no window bounds, \texttt{{[}1575,\ 11575)} & \texttt{10,000} & \texttt{6,967} & \texttt{3,033} & \texttt{30.33\%} \\
Bounded, with window bounds, \texttt{{[}1575,\ 11575)} & \texttt{10,000} & \texttt{9,835} & \texttt{165} & \texttt{1.65\%} \\
Expansion, with window bounds, \texttt{{[}11575,\ 111575)} & \texttt{100,000} & \texttt{93,929} & \texttt{6,071} & \texttt{6.07\%} \\
\end{longtable}
}

The \texttt{20.10\%} calibration rate was run \emph{before} window-bounds were enabled
and is included for reference only; once window-cardinality pruning is on
(see §3.2), the residual sits in the \texttt{1}--\texttt{6\%} band depending on the
chunk-ID range.

\textbf{Recursive refinement of \texttt{UNKNOWN} chunks.} Every chunk timing out at
the broad layer can be refined by extending the witness-pin prefix with the
next-degree split variables and re-solving the resulting subchunks. The number
of emitted rows is not a raw \texttt{2\^{}d} fan-out: AP-prefix pruning eliminates many
descendants before a solver call is made. The retained refinement diagnostics
are shown below. The depth labels follow the L1/L2/L3/L4 refinement ladder of
§2.4.

{\def\LTcaptype{none} 
\begin{longtable}[]{@{}
  >{\raggedright\arraybackslash}p{(\linewidth - 10\tabcolsep) * \real{0.1364}}
  >{\raggedright\arraybackslash}p{(\linewidth - 10\tabcolsep) * \real{0.1364}}
  >{\raggedleft\arraybackslash}p{(\linewidth - 10\tabcolsep) * \real{0.1818}}
  >{\raggedleft\arraybackslash}p{(\linewidth - 10\tabcolsep) * \real{0.1818}}
  >{\raggedleft\arraybackslash}p{(\linewidth - 10\tabcolsep) * \real{0.1818}}
  >{\raggedleft\arraybackslash}p{(\linewidth - 10\tabcolsep) * \real{0.1818}}@{}}
\toprule\noalign{}
\begin{minipage}[b]{\linewidth}\raggedright
Stream
\end{minipage} & \begin{minipage}[b]{\linewidth}\raggedright
Source
\end{minipage} & \begin{minipage}[b]{\linewidth}\raggedleft
Depth added
\end{minipage} & \begin{minipage}[b]{\linewidth}\raggedleft
Rows emitted
\end{minipage} & \begin{minipage}[b]{\linewidth}\raggedleft
INFEASIBLE
\end{minipage} & \begin{minipage}[b]{\linewidth}\raggedleft
UNKNOWN
\end{minipage} \\
\midrule\noalign{}
\endhead
\bottomrule\noalign{}
\endlastfoot
Sample-100 L1 & random sample of \texttt{100} broad UNKs & \texttt{+16} (to depth \texttt{40}) & \texttt{299,375} & \texttt{299,374} & \texttt{1} \\
Sample-100 tail & one Sample-100 residual & \texttt{+8} (to depth \texttt{48}) & \texttt{8} & \texttt{8} & \texttt{0} \\
Sample-500 L1 & stratified sample of \texttt{500} broad UNKs & \texttt{+16} (to depth \texttt{40}) & \texttt{2,076,105} & \texttt{2,076,095} & \texttt{10} \\
Sample-500 tail8 & \texttt{10} Sample-500 residuals & \texttt{+8} (to depth \texttt{48}) & \texttt{76} & \texttt{68} & \texttt{8} \\
Sample-500 level3 & \texttt{8} tail8 residuals & \texttt{+8} (to depth \texttt{56}) & \texttt{8} & \texttt{2} & \texttt{6} \\
Sample-500 level4 & \texttt{6} level3 residuals at \texttt{600}-s wall & \texttt{+8} (to depth \texttt{64}) & \texttt{6} & \texttt{6} & \texttt{0} \\
\end{longtable}
}

The final Sample-500 level4 cleanup closed all \texttt{6} deep residuals within the
\texttt{600}-s cap, with a worst-case solver time of \texttt{599.74} s --- within \texttt{0.26} s of
the cap, so this cleanup is at the practical limit of the refinement
strategy at its current parameter settings. Crucially, none of these
levels produced a \texttt{FEASIBLE} row.

The \texttt{6,071} \texttt{UNKNOWN} chunks of the expansion batch were \emph{not} fully
refined by this campaign. Three subsets of them were sampled for
diagnostic experiments: a uniform random \texttt{100}-chunk recap study
(§3.2 and §4.2), a structural-mining analysis (§4.1), and the HiGHS
attack of §4.3.

\begin{figure}[t]
\centering
\resizebox{0.96\linewidth}{!}{%
\begin{tikzpicture}[
  >=Latex,
  node distance=7mm and 8mm,
  box/.style={
    draw,
    rounded corners=2pt,
    align=center,
    inner xsep=6pt,
    inner ysep=5pt,
    font=\small,
    minimum width=34mm
  },
  smallbox/.style={
    draw,
    rounded corners=2pt,
    align=center,
    inner xsep=5pt,
    inner ysep=4pt,
    font=\footnotesize,
    minimum width=30mm
  },
  edge/.style={->, thick}
]
\node[box] (broad) {100k broad expansion\\93,929 INF / 6,071 UNK};
\node[box, below=of broad] (recap) {100 sampled UNKs\\300s recap};
\node[box, below=of recap] (t1) {T1: 45 survivors};
\node[smallbox, below left=10mm and 17mm of t1] (t1a) {T1a: 25\\HiGHS 8h INF};
\node[smallbox, below right=10mm and 17mm of t1] (t1b) {T1b: 20\\HiGHS 8h UNK\\dual 0.0};
\node[smallbox, below left=10mm and 7mm of t1b] (cdcl) {18\\CDCL UNSAT\\DRAT verified};
\node[smallbox, below right=10mm and 7mm of t1b] (t1c) {T1c: 2\\40959, 48895\\all tested arms UNK};

\draw[edge] (broad) -- node[right, font=\scriptsize] {uniform sample} (recap);
\draw[edge] (recap) -- node[right, font=\scriptsize] {45 remain UNK} (t1);
\draw[edge] (t1) -- node[left, font=\scriptsize] {closed} (t1a);
\draw[edge] (t1) -- node[right, font=\scriptsize] {LP-flat} (t1b);
\draw[edge] (t1b) -- node[left, font=\scriptsize] {CDCL closes} (cdcl);
\draw[edge] (t1b) -- node[right, font=\scriptsize] {survives} (t1c);
\end{tikzpicture}
}
\caption{Funnel from the retained `100k` broad expansion batch to the final
two-chunk T1c residual. T1 is a `45`-chunk subset produced by recapping a
uniform random `100`-chunk sample of the `6,071` broad UNKNOWN chunks at
`300` seconds.}
\label{fig:t1-funnel}
\end{figure}
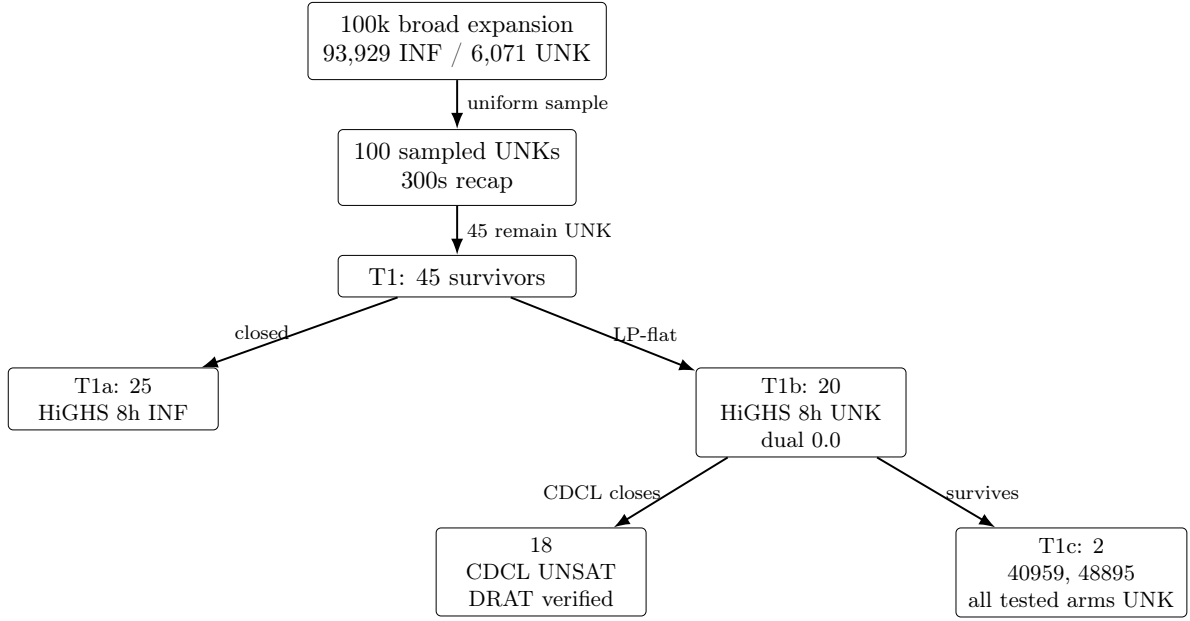

\subsection{3.2 The window-cardinality A/B}\label{the-window-cardinality-ab}

The single most impactful intervention in the campaign was adding
window-cardinality inequalities derived from the OEIS A003002 b-file.
For every window \texttt{{[}a,\ a\ +\ L\ -\ 1{]}\ ⊆\ {[}1,\ 212{]}} and every length \texttt{L} with
\texttt{r\_3(L)\ \textless{}\ min(L,\ K)}, we add \texttt{sum\_\{i\ ∈\ window\}\ x\_i\ ≤\ r\_3(L)}. For
\texttt{N\ =\ 212,\ K\ =\ 44} this generates \texttt{22,154} inequalities, on top of the
\texttt{11,130} 3-AP triple inequalities and the symmetry-breaking lex
constraint.

We measured the effect on three batches of varying size:

{\def\LTcaptype{none} 
\begin{longtable}[]{@{}
  >{\raggedright\arraybackslash}p{(\linewidth - 8\tabcolsep) * \real{0.1579}}
  >{\raggedleft\arraybackslash}p{(\linewidth - 8\tabcolsep) * \real{0.2105}}
  >{\raggedleft\arraybackslash}p{(\linewidth - 8\tabcolsep) * \real{0.2105}}
  >{\raggedleft\arraybackslash}p{(\linewidth - 8\tabcolsep) * \real{0.2105}}
  >{\raggedleft\arraybackslash}p{(\linewidth - 8\tabcolsep) * \real{0.2105}}@{}}
\toprule\noalign{}
\begin{minipage}[b]{\linewidth}\raggedright
Batch
\end{minipage} & \begin{minipage}[b]{\linewidth}\raggedleft
Chunks
\end{minipage} & \begin{minipage}[b]{\linewidth}\raggedleft
UNK without window-bounds
\end{minipage} & \begin{minipage}[b]{\linewidth}\raggedleft
UNK with window-bounds
\end{minipage} & \begin{minipage}[b]{\linewidth}\raggedleft
Reduction
\end{minipage} \\
\midrule\noalign{}
\endhead
\bottomrule\noalign{}
\endlastfoot
\texttt{{[}575,\ 675)} & \texttt{100} & \texttt{27.00\%} & \texttt{0.00\%} & \texttt{−27.0\ pp} \\
\texttt{{[}1575,\ 11575)} & \texttt{10,000} & \texttt{30.33\%} & \texttt{1.65\%} & \texttt{−28.7\ pp} \\
\texttt{{[}11575,\ 111575)} & \texttt{100,000} & n/a & \texttt{6.07\%} & --- \\
\end{longtable}
}

Both controlled A/Bs show a roughly \texttt{28}-percentage-point reduction in
\texttt{UNK} rate at the \texttt{60}-s broad wall cap. The window-bound family is
strictly stronger than the baseline 3-AP family at this problem size. In the
retained logs it also lowered total broad-pass solver time, because many
formerly hard chunks closed early under the extra bounds.

The \texttt{{[}575,\ 675)} A/B is a \texttt{100}-chunk subrange of the calibration range
reported in §3.1; the difference between its \texttt{27.00\%} no-window \texttt{UNK} rate
and the calibration range's \texttt{20.10\%} reflects the structural non-uniformity of
chunk-ID space rather than a change in model configuration.

However, the \texttt{UNK} rate is non-monotone in the chunk-ID range: the small
controlled A/B on \texttt{{[}575,\ 675)} closes to \texttt{0\%}, the \texttt{10k} bounded batch
sits at \texttt{1.65\%}, and the \texttt{100k} expansion batch sits at \texttt{6.07\%}. This
indicates that the chunk-ID space is structurally non-uniform --- later
chunk IDs (which encode different witness-pin assignments) are harder
in a way that window-bounds do not fully compensate for. The bucket
analysis in §4.1 quantifies this non-uniformity.

\subsection{3.3 Solver configuration and engineering}\label{solver-configuration-and-engineering}

All CP-SAT calls use the following configuration unless explicitly
overridden in an experiment:

\begin{itemize}
\tightlist
\item
  Model: decision-mode \texttt{sum\_i\ x\_i\ =\ 44}, \texttt{11,130} 3-AP linear
  inequalities, \texttt{22,154} window-cardinality inequalities, lex
  reflection symmetry break, endpoint forcing \texttt{x\_1\ =\ x\_\{212\}\ =\ 1}.
\item
  Branch strategy: variable selection by AP-incidence degree, value
  selection \texttt{min}, \texttt{fixed\_search} to disable CP-SAT's portfolio.
\item
  Solver: OR-Tools CP-SAT \cite{or-tools}, \texttt{8} workers per task, fixed solver seed
  and fixed search parameters, wall cap as noted (most commonly \texttt{60}-s
  for broad, \texttt{300}-s and \texttt{600}-s for recap experiments).
\end{itemize}

The SLURM emitter, shard collector, tail emitter, and proof-state manager
implement the workload pipeline. Output is
written as line-delimited JSON with one record per chunk; shards are
written to a temporary path and atomically renamed on completion so
that partial output from killed tasks cannot corrupt downstream
collection. The full campaign is reproducible end-to-end via the
\texttt{unity\_handoff.sh} driver.

\subsection{3.4 Zero FEASIBLE across the campaign}\label{zero-feasible-across-the-campaign}

Aggregating across the broad pass, the refinement loop, the recap
studies, the lever experiments of §3.5, and the HiGHS attack of §4.3,
the campaign solved millions of \texttt{(N,\ K,\ fixed\_in,\ fixed\_out)} CP-SAT
subproblems, including prefix-closure work not tabulated above, \texttt{45}
HiGHS MIP subproblems, and the CDCL/SAT and proof-producing reruns reported
in §4.3. The number of
subproblems returning \texttt{FEASIBLE} is \textbf{zero}.

The \texttt{FEASIBLE} count is the only signal that would directly disprove
\texttt{r\_3(212)\ ≤\ 43}; its absence does not constitute a proof of the upper
bound, but it is strong empirical support, especially given that the
campaign forced the endpoints \texttt{1} and \texttt{212}, which any 44-set must contain by
the known value \texttt{r\_3(211)\ =\ 43}, and then explored a depth-\texttt{24} prefix of the
verified 43-witness in the processed chunk ranges. Within the processed
expansion range, a 44-element 3-AP-free subset of \texttt{{[}1,\ 212{]}}, if one existed,
would have to lie in the \texttt{6,071} unresolved broad \texttt{UNKNOWN} chunks (or in
their deeper refinements). Globally, the much larger unprocessed remainder of
the full depth-\texttt{24} sweep remains open as well.

\subsection{3.5 Search-tuning levers and their effect sizes}\label{search-tuning-levers-and-their-effect-sizes}

We tested five CP-SAT-side search-tuning levers. The
full lever inventory, including the HiGHS substitution and the
pair-AND Tseitin experiment, is consolidated in §4.4; here we report
only the three CP-SAT-side levers that operate at the broad layer.

{\def\LTcaptype{none} 
\begin{longtable}[]{@{}
  >{\raggedright\arraybackslash}p{(\linewidth - 8\tabcolsep) * \real{0.1765}}
  >{\raggedright\arraybackslash}p{(\linewidth - 8\tabcolsep) * \real{0.1765}}
  >{\raggedright\arraybackslash}p{(\linewidth - 8\tabcolsep) * \real{0.1765}}
  >{\raggedleft\arraybackslash}p{(\linewidth - 8\tabcolsep) * \real{0.2353}}
  >{\raggedleft\arraybackslash}p{(\linewidth - 8\tabcolsep) * \real{0.2353}}@{}}
\toprule\noalign{}
\begin{minipage}[b]{\linewidth}\raggedright
Lever
\end{minipage} & \begin{minipage}[b]{\linewidth}\raggedright
Mechanism
\end{minipage} & \begin{minipage}[b]{\linewidth}\raggedright
Bucket
\end{minipage} & \begin{minipage}[b]{\linewidth}\raggedleft
Baseline UNK
\end{minipage} & \begin{minipage}[b]{\linewidth}\raggedleft
Treatment UNK
\end{minipage} \\
\midrule\noalign{}
\endhead
\bottomrule\noalign{}
\endlastfoot
Split-vars reorder & Permute the depth-\texttt{24} prefix so hot pins (values \texttt{\{68,\ 75,\ 70,\ 76,\ 91\}}) lead & \texttt{{[}61575,\ 66575)} & \texttt{13.60\%} & \texttt{12.96\%} \\
Wall-cap extension \texttt{60s\ →\ 300s} & Same model, same prefix, longer per-chunk budget & random \texttt{100} from expansion \texttt{UNK} & \texttt{100.00\%} & \texttt{45.00\%} \\
Recursive deepening & Refine \texttt{UNK} to depth \texttt{64} at \texttt{600}-s cap & \texttt{6} L3 residuals & \texttt{100.00\%} & \texttt{0.00\%} \\
\end{longtable}
}

The reorder lever was tested for completeness but is expected to be a
null result on principle: the depth-\texttt{24} broad split iterates over the
\texttt{2\^{}24} IN/OUT prefix assignments, and within-prefix variable ordering
does not change the set of surviving chunks under AP-prefix pruning
--- only the order in which they are emitted to SLURM. The within-noise
result confirms this.

The wall-cap extension lever moves the residual substantially on the
first doubling but exhibits a saturating closure curve under further
extension. The \texttt{60s\ →\ 120s\ →\ 300s} series is analyzed in §4.2; in
brief, the marginal close-rate per additional second of wall drops
sharply past \texttt{300}-s, which is why the campaign did not pursue
\texttt{600}-s or \texttt{1200}-s broad-layer reruns on the full \texttt{6,071}-chunk
expansion residual.

The recursive-deepening lever, applied locally to small residual sets,
reliably closes individual deep \texttt{UNK} chunks (\texttt{6\ /\ 6} at the L4 level
in §3.1). But the levers it implies --- picking the next-degree split
variables and re-solving up to \texttt{2\^{}8} descendant assignments at a longer wall ---
do not
generalize cheaply to the full \texttt{6,071}-chunk expansion residual. A
naïve depth-\texttt{32} rerun on every expansion \texttt{UNK} would emit roughly
\texttt{1.5} million subproblems and require an order of magnitude more CPU
time than the original broad pass, with no guarantee of closing the
hard core analyzed in §4.

\subsection{3.6 Compute budget}\label{compute-budget}

The retained logs provide measured solver wall time for the main diagnostics.
Since CP-SAT tasks used \texttt{8} workers, the worker-hour estimates below multiply
recorded solver-wall seconds by \texttt{8}; for HiGHS, the analogous estimate is
\texttt{8} threads per one-hour task. These are retained-log estimates, not exact
SLURM accounting totals.

{\def\LTcaptype{none} 
\begin{longtable}[]{@{}
  >{\raggedright\arraybackslash}p{(\linewidth - 4\tabcolsep) * \real{0.3000}}
  >{\raggedleft\arraybackslash}p{(\linewidth - 4\tabcolsep) * \real{0.4000}}
  >{\raggedright\arraybackslash}p{(\linewidth - 4\tabcolsep) * \real{0.3000}}@{}}
\toprule\noalign{}
\begin{minipage}[b]{\linewidth}\raggedright
Layer
\end{minipage} & \begin{minipage}[b]{\linewidth}\raggedleft
Worker-hours
\end{minipage} & \begin{minipage}[b]{\linewidth}\raggedright
Notes
\end{minipage} \\
\midrule\noalign{}
\endhead
\bottomrule\noalign{}
\endlastfoot
Retained broad logs & \texttt{\textasciitilde{}2,276} & includes the \texttt{10k} no-window run, the \texttt{10k} window run, the \texttt{100k} window run, and the \texttt{{[}575,675)} A/B \\
Sample-100 refinement & \texttt{\textasciitilde{}375} & depth-\texttt{40} plus one depth-\texttt{48} tail \\
Sample-500 refinement & \texttt{\textasciitilde{}1,978} & depth-\texttt{40}, depth-\texttt{48}, depth-\texttt{56}, and depth-\texttt{64} cleanup \\
Recap and worst-bucket A/Bs & \texttt{\textasciitilde{}598} & reorder, \texttt{300s} walltime, \texttt{120s} recap, \texttt{300s} recap \\
HiGHS attack on \texttt{45} survivors & \texttt{\textasciitilde{}360} & \texttt{1}-h wall, \texttt{8} threads per task \\
Full T1 HiGHS long-wall audit & \texttt{\textasciitilde{}2,002} & \texttt{45} T1 chunks, \texttt{8}-h cap, job \texttt{58782313}; estimated from retained \texttt{seconds} fields \\
T1b CDCL first run & \texttt{\textasciitilde{}19.8} & \texttt{20} single-core tasks, retained JSON row time \texttt{71,216.85} seconds \\
T1b proof-producing rerun + T1c diagnostic & \texttt{\textasciitilde{}120} & proof emission + 4-cell T1c grid \\
drat-trim verification & \texttt{\textasciitilde{}61} & jobs \texttt{58952708}, \texttt{59058393}, and \texttt{59383874}; all \texttt{18} CDCL-resolved \texttt{T1b\ ∖\ T1c} chunks VERIFIED \\
Total retained logs & \texttt{\textasciitilde{}7,850} worker-hours & excludes interactive debugging and logs not retained locally \\
\end{longtable}
}

The campaign fits comfortably within the \texttt{pi\_ergezerm\_wit\_edu} SLURM
allocation on the \texttt{cpu} partition. The dominant retained costs are the
\texttt{100k} broad expansion, the sample-500 refinement, and the full-T1 long-wall
audit; the CDCL and certificate diagnostics are comparatively cheap. We note
for context that
a full depth-\texttt{24} sweep of the \texttt{12,582,912} AP-pruned chunks of the
witness-split lattice would require orders of magnitude more worker-hours
under the current architecture. More importantly, it is not justified by the
present evidence: the hard pocket of §4 is the binding obstacle, not simply
broad-layer throughput.

\section{4. The structural hard pocket}\label{the-structural-hard-pocket}

The headline empirical finding of the campaign is not the absence of a 44-element
3-AP-free subset of \texttt{{[}1,\ 212{]}} --- which we expected from the OEIS A003002
extrapolation \texttt{r\_3(212)\ =\ 43} --- but the existence of a small, robust subset of
broad subproblems that resist every solver paradigm and search lever we threw
at them. We refer to this subset as the \emph{hard pocket}. This section
characterizes it empirically, summarizes the unsuccessful interventions, and
states the resulting open problem.

\subsection{4.1 Empirical signature on the broad population}\label{empirical-signature-on-the-broad-population}

The \texttt{100,000}-chunk window-bound broad batch over chunk-id range
\texttt{{[}11575,\ 111575)} produced \texttt{93,929} \texttt{INFEASIBLE} chunks, \texttt{6,071} \texttt{UNKNOWN}
chunks, and \texttt{0} \texttt{FEASIBLE} chunks under a \texttt{60}-second per-chunk wall cap. The
\texttt{6.07\%} UNKNOWN rate is non-uniform in two distinct senses.

\textbf{Bucket non-uniformity.} Partitioned into twenty \texttt{5,000}-chunk buckets, the
UNKNOWN rate varies from \texttt{1.64\%} to \texttt{13.60\%}, with the worst bucket
\texttt{{[}61575,\ 66575)} at \texttt{13.60\%}. The longest contiguous UNKNOWN run found in this
batch was length \texttt{27} at chunk IDs \texttt{98277..98303}; many other runs have length
around \texttt{15}. These contiguous runs are a strong indication that UNKNOWN status
is driven by joint structural properties of the depth-\texttt{24} witness-pin
assignment, not by stochastic solver behavior.

\textbf{Pin-OUT enrichment.} For each of the \texttt{24} broad-split witness variables we
computed the conditional UNKNOWN rate given that variable's IN/OUT assignment
in the chunk. Five variables --- values \texttt{68}, \texttt{75}, \texttt{70}, \texttt{76}, \texttt{91}, all in the
dense middle cluster \texttt{{[}67,\ 91{]}} of the verified \texttt{43}-witness --- show a strong
asymmetry: their conditional UNKNOWN rate is roughly \texttt{2.4\%} when pinned IN
and \texttt{9.7\%} when pinned OUT (Fisher log-odds \texttt{+3.5}). The signal is consistent
across the five values. We refer to a chunk in which all five hot pins are
forced OUT as a \emph{middle-out chunk}. Middle-out chunks account for a
disproportionate share of the UNKNOWN tail of the broad batch.

\subsection{4.2 The 300s-resistant tail does not collapse}\label{the-300s-resistant-tail-does-not-collapse}

To test whether the broad-pass UNKNOWNs were merely ``slow'' rather than
structurally hard, we ran two wall-cap recalibration experiments on a uniform
random sample of \texttt{100} UNKNOWN chunks drawn from the \texttt{6,071} baseline. The
results show a saturating, not exponential, closure curve:

{\def\LTcaptype{none} 
\begin{longtable}[]{@{}rrrr@{}}
\toprule\noalign{}
Wall cap & INFEASIBLE & UNKNOWN & Close rate over baseline \\
\midrule\noalign{}
\endhead
\bottomrule\noalign{}
\endlastfoot
60 s & 0 & 100 & 0\% \\
120 s & 31 & 69 & 31\% \\
300 s & 55 & 45 & 55\% \\
\end{longtable}
}

Going from \texttt{60s} to \texttt{120s} (2x) closed \texttt{31} chunks. Going from \texttt{120s} to
\texttt{300s} (2.5x further) closed only \texttt{24} more. Naïve extrapolation to \texttt{600s}
predicts perhaps \texttt{10}--\texttt{15} additional closures. The hard tail does not vanish
under more time.

We applied the same structural mining procedure (single-pin, pair, and triple
log-odds enrichment relative to a matched \texttt{INFEASIBLE} sample) to the \texttt{45}
\texttt{UNKNOWN} chunks that survived the \texttt{300s} cap. The broad pin-OUT signature
weakens substantially. The best high-coverage pair, \texttt{{[}91=OUT,\ 48=OUT{]}}, covers
\texttt{66.67\%} of the survivors with log-odds \texttt{+1.226} --- a modest effect size, far
from the \texttt{+3.5} signal seen on the \texttt{6,071}-row population. Top triples have
stronger log-odds but each covers only a handful of cases. Hamming-distance
clustering of survivor \texttt{fixed\_in} sets shows multiple small clusters rather
than a single dominant niche. We interpret this as evidence that the
\texttt{300s}-resistant subproblems share moderate pin-OUT structure but spread
across many distinct sub-pockets of the depth-\texttt{24} assignment lattice.

\subsection{4.3 LP-paradigm failure and the CDCL break}\label{lp-paradigm-failure-and-the-cdcl-break}

The 300s-resistant pocket might in principle be a CP-SAT-specific
phenomenon --- an artifact of constraint propagation's inability to exploit
implicit LP structure. To test this, we re-attacked all \texttt{45} survivors with
the HiGHS open-source MIP solver \cite{huangfu-hall-2018-highs}, which combines branch-and-bound, LP
relaxation, presolve, cutting planes, and primal heuristics. We used a
one-hour wall cap per chunk, \texttt{8} threads per chunk, and the identical
constraint set
(decision-form \texttt{sum\ x\_i\ =\ 44}, all \texttt{11,130} 3-AP triple inequalities, all
\texttt{22,154} window-cardinality inequalities, plus the broad chunk's \texttt{fixed\_in}
and \texttt{fixed\_out} assignments tightened into variable bounds).

The one-hour result was negative:

The two rows in the following table are not a common-sample comparison: the
HiGHS attack targets precisely the \texttt{45} chunks that survived the \texttt{300}-s
CP-SAT recap, i.e.~the harder subset of the \texttt{100} recap inputs.

{\def\LTcaptype{none} 
\begin{longtable}[]{@{}
  >{\raggedright\arraybackslash}p{(\linewidth - 10\tabcolsep) * \real{0.1304}}
  >{\raggedleft\arraybackslash}p{(\linewidth - 10\tabcolsep) * \real{0.1739}}
  >{\raggedleft\arraybackslash}p{(\linewidth - 10\tabcolsep) * \real{0.1739}}
  >{\raggedleft\arraybackslash}p{(\linewidth - 10\tabcolsep) * \real{0.1739}}
  >{\raggedleft\arraybackslash}p{(\linewidth - 10\tabcolsep) * \real{0.1739}}
  >{\raggedleft\arraybackslash}p{(\linewidth - 10\tabcolsep) * \real{0.1739}}@{}}
\toprule\noalign{}
\begin{minipage}[b]{\linewidth}\raggedright
Solver
\end{minipage} & \begin{minipage}[b]{\linewidth}\raggedleft
Constraints
\end{minipage} & \begin{minipage}[b]{\linewidth}\raggedleft
Wall cap
\end{minipage} & \begin{minipage}[b]{\linewidth}\raggedleft
Closed (INF)
\end{minipage} & \begin{minipage}[b]{\linewidth}\raggedleft
UNKNOWN
\end{minipage} & \begin{minipage}[b]{\linewidth}\raggedleft
Aggregate solver wall time
\end{minipage} \\
\midrule\noalign{}
\endhead
\bottomrule\noalign{}
\endlastfoot
CP-SAT, window-bounds & 3-AP + window + symmetry & 300 s & 55 / 100 & 45 / 100 & bounded by \textasciitilde8.3 wall-h \\
HiGHS, window-bounds & 3-AP + window + endpoint & 3,600 s & \textbf{0 / 45} & 45 / 45 & 45.0 wall-h \\
\end{longtable}
}

Across \texttt{162,003} solver-seconds and \texttt{3,181,316} explored MIP nodes, HiGHS did
not close a single chunk. No \texttt{FEASIBLE} row appeared. The recorded HiGHS dual
bound stayed at its uninformative default value in every task, which is itself
evidence that the LP-relaxation route was not discovering a useful certificate
for these instances.

We then re-attacked all \texttt{45} T1 chunks under an extended \texttt{8}-hour HiGHS wall,
with LP progress logging enabled. The result is mixed: \texttt{25\ /\ 45} chunks closed
\texttt{INFEASIBLE}, while \texttt{20\ /\ 45} returned \texttt{UNKNOWN} at the full cap with dual
bound still pinned at \texttt{0.0}. We call this \texttt{20}-chunk LP-paradigm-resistant
subset \textbf{T1b}. The audit consumed \texttt{901,073} solver-seconds and \texttt{25,196,448}
MIP nodes.

To test whether T1b is invariant under solver architecture more broadly, we
re-attacked it with a CDCL/SAT solver (CaDiCaL via PySAT
\cite{biere-cadical,pysat-2018}, single-threaded,
\texttt{4}-hour wall, encoding restricted to 3-AP triples + cardinality + chunk pins;
no window-cardinality clauses). CDCL closed \texttt{18\ /\ 20} chunks \texttt{UNSAT}. We call
the surviving \texttt{2}-chunk residual \textbf{T1c} = \texttt{\{40959,\ 48895\}}. A T1c diagnostic
at extended wall (\texttt{12} h pure CDCL) and with totalizer-encoded window
constraints for lengths \texttt{\{31,\ 100,\ 199\}} (\texttt{4} h) also returned \texttt{UNKNOWN} on
both chunks, so T1c is resistant to all tested paradigms in this campaign.

A proof-producing rerun of the \texttt{18} CDCL-UNSAT chunks initially emitted DRAT
proofs for \texttt{17} chunks; the remaining chunk, \texttt{32735}, required a larger-memory
proof rerun and then also emitted a DRAT proof. Independent \texttt{drat-trim}
verification confirmed all \texttt{18\ /\ 18} emitted proofs. The final two certificates
were the slowest: \texttt{63231} verified in \texttt{49,758.11} seconds and \texttt{32735} verified
in \texttt{80,960.68} seconds under the final \texttt{24}-hour verifier pass
\cite{heule-hunt-biere-drat-2014,cruz-filipe-lrat-2017}.

The refined paradigm-invariance picture is therefore: \textbf{LP-paradigm methods
(CP-SAT constraint propagation and HiGHS LP-relaxation MIP) fail uniformly on
T1b}, and the CDCL clause-learning paradigm closes \texttt{18\ /\ 20} of those chunks,
all of which are independently verified. The genuinely paradigm-invariant
residual is T1c, of size \texttt{2}.

\subsection{4.4 Levers tested and their outcomes}\label{levers-tested-and-their-outcomes}

For completeness, we record every search-tuning lever we tested on the hard
bucket \texttt{{[}61575,\ 66575)} or the \texttt{300s}-resistant subset. The CP-SAT/MIP-side
tuning levers did not remove the hard pocket; the one qualitative exception is
the CDCL paradigm switch, which closes most of T1b and therefore narrows, rather
than merely tunes, the residual.

{\def\LTcaptype{none} 
\begin{longtable}[]{@{}
  >{\raggedright\arraybackslash}p{(\linewidth - 4\tabcolsep) * \real{0.3333}}
  >{\raggedright\arraybackslash}p{(\linewidth - 4\tabcolsep) * \real{0.3333}}
  >{\raggedright\arraybackslash}p{(\linewidth - 4\tabcolsep) * \real{0.3333}}@{}}
\toprule\noalign{}
\begin{minipage}[b]{\linewidth}\raggedright
Lever
\end{minipage} & \begin{minipage}[b]{\linewidth}\raggedright
Mechanism
\end{minipage} & \begin{minipage}[b]{\linewidth}\raggedright
Result
\end{minipage} \\
\midrule\noalign{}
\endhead
\bottomrule\noalign{}
\endlastfoot
Window-cardinality (OEIS A003002) & Add \texttt{sum\ x\ ≤\ r\_3(L)} for each window & UNK rate on full \texttt{10k} batch: \texttt{30.33\%\ →\ 1.65\%}. Strong but plateaus. \\
Split-vars reorder (hot pins first) & Permute the depth-\texttt{24} split prefix & \texttt{13.60\%\ →\ 12.96\%} on the worst bucket. Noise. \\
Wall-cap extension & \texttt{60s\ →\ 300s} on the broad pass & Saturating curve, see §4.2. \\
Targeted pair-AND Tseitin propagators & Explicit \texttt{pair{[}a,c{]}\ =\ x{[}a{]}∧x{[}c{]}} BoolVars on midpoints in \texttt{{[}67,\ 91{]}} & Control \texttt{67.84s}, treatment \texttt{71.75s} on a sampled residual. No effect. \\
Recursive deepening & Refinement at depths \texttt{40}, \texttt{48}, \texttt{56} & All \texttt{500} stratified-sample base chunks close, but with a non-trivial tail; the final six level-3 residuals needed up to \texttt{599.74s} at the \texttt{600s} cap. \\
HiGHS MIP with LP/cut machinery & Replace CP-SAT entirely & \texttt{0\ /\ 45} closed at \texttt{1} hour; \texttt{25\ /\ 45} closed in the full \texttt{8}-hour audit, while \texttt{20\ /\ 45} retained dual bound \texttt{0.0}. \\
Pure CDCL/SAT & Encode T1b as CNF with 3-AP clauses + cardinality + pins, no windows & \texttt{18\ /\ 20} HiGHS-flat chunks closed \texttt{UNSAT} in 4 hours; \texttt{2\ /\ 20} remained \texttt{UNKNOWN}; no \texttt{SAT} rows. This breaks the strong solver-invariance framing. \\
\end{longtable}
}

The pair-AND result is worth a closer note. The added constraints are
mathematically redundant with the existing 3-AP triple inequalities --- they
encode the same forbidden configurations, just with an explicit Tseitin
variable per pair. The hope was that this would give CP-SAT more
``propagation hooks'' in the structural pocket. The negative result is
consistent with the broader picture: the pocket's hardness is not an
encoding issue, it is a search-space issue.

\subsection{4.5 A conjecture about the pocket}\label{a-conjecture-about-the-pocket}

We end §4 with a working conjecture about T1c.

CDCL closed \texttt{18\ /\ 20} of T1b, refuting the strongest reading of
solver-paradigm invariance for the full \texttt{20}-chunk LP-flat subset. The
conjecture below is deliberately narrower: it concerns only the two chunks
that survived the CDCL break.

Each T1c subproblem corresponds to a depth-\texttt{24} fixed assignment in which
\texttt{(i)} the LP relaxation upper bound on \texttt{sum\ x\_i} is at most one above the
decision threshold \texttt{K\ =\ 44}, leaving no room for LP-based cuts to prove
infeasibility, and \texttt{(ii)} the integer infeasibility certificate is too
combinatorially diffuse to fit in the working memory of current CDCL
clause-learning solvers under a \texttt{12}-hour wall. The audit of §4.3 bounds the
T1c population at \texttt{2} chunks within the \texttt{100,000}-chunk window-bound expansion
residual; we do not have a population-level estimate beyond the audited
region.

If this picture is correct, closing T1c requires either (a) a
problem-specific additive-combinatorial upper bound tighter than the OEIS
window-cardinality family, or (b) a custom branch-and-bound or proof-search
system that exploits problem structure neither current general-purpose solver
paradigm captures. Both are framed as open computational problems in §5.

\subsection{\texorpdfstring{4.6 What this means for \texttt{r\_3(212)}}{4.6 What this means for r\_3(212)}}\label{what-this-means-for-r_3212}

The combined CP-SAT, HiGHS, and CDCL evidence --- \texttt{0} \texttt{FEASIBLE}/\texttt{SAT} rows
across millions of subproblems and all audited hard-pocket diagnostics ---
strongly supports the OEIS-conjectured value \texttt{r\_3(212)\ =\ 43}, but the
surviving T1c pocket means we do not have a formal proof of the upper bound
\texttt{r\_3(212)\ ≤\ 43}. The \texttt{6,071} baseline UNKNOWN chunks (and the much larger
unobserved tail of the full \texttt{12,582,912}-chunk depth-\texttt{24} sweep) remain an
open obstruction. The lower bound \texttt{r\_3(212)\ ≥\ 43} from the verified witness in
§2 is unaffected. The trivial upper bound \texttt{r\_3(212)\ ≤\ 44} follows from
monotonicity and OEIS \texttt{r\_3(211)\ =\ 43}, so the gap between our certified bounds
is exactly one.

Closing that final unit gap would require resolving T1c and independently
certifying the entire depth-\texttt{24} sweep, including the unprocessed remainder of
the \texttt{12,582,912}-chunk T3 lattice. T1c is the sharpest audited open instance,
not necessarily a sufficient stepping stone. We release the hard-pocket
benchmark instances in §5 so specialists can attempt this directly.

\section{5. Open problem and benchmark release}\label{open-problem-and-benchmark-release}

The campaign establishes strong empirical support for \texttt{r\_3(212)\ =\ 43}
(§3.4) but leaves a hard pocket that resists both CP-SAT and HiGHS
under every lever we tested (§4). To turn that pocket into a tractable
research target rather than an open compute frontier, this section
states the residual problem cleanly and releases the surviving
instances as a tiered benchmark.

\subsection{5.1 Statement of the open problem}\label{statement-of-the-open-problem}

\begin{quote}
\textbf{Problem (R3-212-UB).} Let \texttt{A\ ⊆\ {[}1,\ 212{]}} with \texttt{\textbar{}A\textbar{}\ =\ 44}. Determine
whether there exists \texttt{A} containing no nontrivial 3-term arithmetic
progression. The verified \texttt{43}-element witness in
\texttt{results/N212\_K43\_witness.json} (§2.1) and the value \texttt{r\_3(211)\ =\ 43}
from OEIS A003002 jointly imply that any such \texttt{A} must contain both
endpoints \texttt{1} and \texttt{212}. Equivalently, either exhibit one such \texttt{A}
or certify the infeasibility of the decision problem
\texttt{sum\_i\ x\_i\ =\ 44} over the \texttt{3}-AP constraint family on \texttt{{[}1,\ 212{]}}
with \texttt{x\_1\ =\ x\_\{212\}\ =\ 1}.
\end{quote}

A solution to R3-212-UB resolves the unit gap \texttt{r\_3(212)\ ∈\ \textbackslash{}\{43,\ 44\textbackslash{}\}}
in either direction. Our multi-million-subproblem CP-SAT campaign and
the \texttt{45}-instance HiGHS attack returned \texttt{0} \texttt{FEASIBLE} rows, which is
evidence for the infeasibility branch but not a proof. The
benchmark instances released below pinpoint the subproblems on which
generic combinatorial-optimization solvers currently fail.

\subsection{5.2 Benchmark instance tiers}\label{benchmark-instance-tiers}

We release three tiered benchmark sets. Each JSONL row records the chunk ID,
fixed assignments, witness-pin prefix, solver status, and timing data; the
common constraint family (3-AP triples, endpoint forcing, and OEIS window
bounds) is reconstructed by the repository scripts. The tiers form a ladder
from the smallest solver-resistant pocket to the full upper-bound proof.

{\def\LTcaptype{none} 
\begin{longtable}[]{@{}
  >{\raggedright\arraybackslash}p{(\linewidth - 8\tabcolsep) * \real{0.1875}}
  >{\raggedright\arraybackslash}p{(\linewidth - 8\tabcolsep) * \real{0.1875}}
  >{\raggedleft\arraybackslash}p{(\linewidth - 8\tabcolsep) * \real{0.2500}}
  >{\raggedright\arraybackslash}p{(\linewidth - 8\tabcolsep) * \real{0.1875}}
  >{\raggedright\arraybackslash}p{(\linewidth - 8\tabcolsep) * \real{0.1875}}@{}}
\toprule\noalign{}
\begin{minipage}[b]{\linewidth}\raggedright
Tier
\end{minipage} & \begin{minipage}[b]{\linewidth}\raggedright
Artifact
\end{minipage} & \begin{minipage}[b]{\linewidth}\raggedleft
Size
\end{minipage} & \begin{minipage}[b]{\linewidth}\raggedright
Resistance level
\end{minipage} & \begin{minipage}[b]{\linewidth}\raggedright
Recommended target
\end{minipage} \\
\midrule\noalign{}
\endhead
\bottomrule\noalign{}
\endlastfoot
T1a & T1a JSONL & \texttt{25} chunks & closed by HiGHS at \texttt{8}h (dual = \texttt{inf}) & reference / regression test \\
T1b ∖ T1c & T1b-minus-T1c JSONL & \texttt{18} chunks & LP-paradigm-resistant; closed by CDCL; all \texttt{18\ /\ 18} emitted DRAT proofs verified by \texttt{drat-trim} & certified CDCL benchmark \\
T1c & T1c JSONL & \texttt{2} chunks & resistant to CP-SAT, HiGHS LP-MIP, pure CDCL @ \texttt{12}h, and windowed CDCL @ \texttt{4}h & minimum-viable proof step \\
T2 & T2 JSONL & \texttt{6,071} chunks & survived \texttt{60}-s CP-SAT broad pass with window bounds & full closure of the expansion residual \\
T3 & deterministic generator & \texttt{12,582,912} chunks & unprocessed remainder of the witness-split lattice & full upper-bound proof \\
\end{longtable}
}

The concrete filenames are \path{results/N212_K44_t1a25.jsonl},
\path{results/N212_K44_t1b_minus_t1c.jsonl},
\path{results/N212_K44_t1c2.jsonl}, and
\path{results/N212_K44_window100k_unknowns.jsonl}.

Tier T1c is the campaign's sharpest open problem: \texttt{2} chunks resistant to
every tested solver paradigm under generous wall caps. A successful T1c closure
either disproves \texttt{r\_3(212)\ ≤\ 43} (one \texttt{FEASIBLE}/\texttt{SAT} row suffices, after
witness verification) or eliminates the audited four-paradigm-resistant
residual, leaving the unit gap depending on the unprocessed remainder of T3.

Tier T2 is the canonical ``close the campaign at the broad layer''
target: closing all \texttt{6,071} chunks of the \texttt{100k} expansion batch is
necessary, though not sufficient, for a full upper-bound proof.

Tier T3 is the full upper-bound certificate: closing the entire
depth-\texttt{24} AP-pruned sweep. The instance generator (§2.2) emits the
required chunks deterministically from the witness file and the
OEIS b-file; the storage cost of the full sweep is dominated by
output rather than input.

\subsection{5.3 Minimum-viable proof requirements}\label{minimum-viable-proof-requirements}

A formal proof of \texttt{r\_3(212)\ ≤\ 43} from the released benchmark requires:

\begin{enumerate}
\def\labelenumi{\arabic{enumi}.}
\tightlist
\item
  \textbf{Closure of tier T3.} Every chunk in the depth-\texttt{24} AP-pruned
  sweep must return \texttt{INFEASIBLE} under a verified solver, or one
  chunk must return a \texttt{FEASIBLE} \texttt{44}-element witness.
\item
  \textbf{Solver verification.} We recommend that any solver used for closure
  either produce machine-checkable proof certificates (DRAT, LRAT, or
  equivalent), or be independently reproduced under a second solver paradigm.
  This is a verification target rather than a condition already met by every
  row of the present campaign; §6.2 records the remaining certificate gaps.
\item
  \textbf{Constraint-set verification.} The \texttt{11,130} 3-AP triple
  inequalities, the \texttt{22,154} window-cardinality inequalities, and
  the endpoint forcing must be checked against the formal
  definition of \texttt{r\_3(N)}. The verifier \texttt{r3\_verify.py} performs
  this check on any candidate witness; an analogous check for the
  constraint generation is included in the repository.
\end{enumerate}

The minimum-viable result short of a full proof is closure of T1c
under a solver paradigm that produces a usable bound certificate
(e.g., LP dual values that improve under cuts, or a CDCL-style
unsatisfiability proof on a SAT encoding of the same constraints).
This would establish that the hard pocket is not solver-architecture
invariant after all, contradicting the conjecture of §4.5.

\subsection{5.4 Approaches we could not test}\label{approaches-we-could-not-test}

The following techniques are plausibly stronger than CP-SAT and HiGHS
on this pocket but were outside the scope of our campaign. We flag
them as natural follow-on directions:

\begin{itemize}
\tightlist
\item
  \textbf{Fourier-analytic upper bounds.} Behrend-style constructions
  achieve \texttt{r\_3(N)\ =\ N\ ·\ exp(-c·sqrt(log\ N))} lower bounds; a
  matching Fourier-analytic upper bound for small \texttt{N} would give a
  certificate independent of combinatorial search. The Bloom--Sisask
  framework for \texttt{r\_3(N)\ =\ O(N\ /\ (log\ N)\^{}\{1+c\})} is asymptotic but
  the underlying density-increment machinery may yield finite-\texttt{N}
  bounds tighter than the OEIS window-cardinality family.
\item
  \textbf{Multi-window partition bounds.} OEIS A003002 is used here as
  a single-window family \texttt{sum\_\{{[}a,\ a+L)\}\ x\_i\ ≤\ r\_3(L)}. A
  partition-cardinality bound across multiple disjoint windows of
  varying length might cut the LP relaxation more aggressively
  than any single-window family.
\item
  \textbf{SAT with proof logging.} A pure CNF encoding has already closed
  \texttt{18\ /\ 20} T1b chunks using PySAT/CaDiCaL without OEIS window constraints.
  Follow-up proof-producing runs emitted DRAT proofs for all \texttt{18} CDCL-closed
  chunks, and \texttt{drat-trim} independently verified all \texttt{18\ /\ 18}. The research
  target is now T1c: try longer CDCL walls, different cardinality encodings,
  full-window encodings, or native proof-producing SAT solvers on chunks
  \texttt{40959} and \texttt{48895}.
\item
  \textbf{Lean/formal-proof-search benchmark.} The repository now includes a compact
  Lean 4 / Mathlib 4 formalization \cite{lean4-2021,mathlib4} under \texttt{lean/}: shared definitions
  (\texttt{R3Base.lean}), the verified \texttt{43}-witness statement
  (\texttt{R3\_212\_Witness.lean}), and two AlphaProof-style T1c targets
  (\texttt{R3\_T1c\_40959.lean}, \texttt{R3\_T1c\_48895.lean}) with the expected answer \texttt{false}.
  These files are intended as a starting point for an A003002 / \texttt{r\_3(N)} entry
  in formal-conjecture repositories \cite{formal-conjectures}. Recent AlphaProof Nexus-style workflows
  have resolved Erdős problems of comparable complexity at low per-problem cost
  when a Lean target exists \cite{tsoukalas-alphaproof-nexus-2026}, so T1c is well-shaped for that paradigm.
\item
  \textbf{Custom branch-and-bound.} A solver that exploits problem-
  specific symmetries and dominance relations --- for instance,
  reflection symmetry after endpoint forcing, or domination of one
  window-bound by another at specific fixed-pin configurations ---
  could prune branches that neither CP-SAT nor HiGHS recognize as
  redundant.
\item
  \textbf{Symmetric Difference Encoding / Lasserre hierarchy at low
  degree.} A degree-\texttt{4} or degree-\texttt{6} SDP relaxation of the
  3-AP-free constraint might separate the surviving T1 instances
  from feasibility where the LP relaxation cannot.
\end{itemize}

We do not advocate for any single direction. The benchmark release
is intended to let specialists pick the technique closest to their
toolkit.

\subsection{5.5 Reproducibility and release}\label{reproducibility-and-release}

The campaign code, configuration, and result logs are released at
the repository accompanying this preprint. Specifically:

\begin{itemize}
\tightlist
\item
  All Python sources implementing the architecture of §2 are present
  with module-level documentation.
\item
  The SLURM scripts (\texttt{submit\_*.sbatch}) and the driver
  \texttt{unity\_handoff.sh} reproduce the full campaign on any
  SLURM-compatible cluster with OR-Tools and \texttt{highspy} installed.
\item
  The verified \texttt{43}-witness, the OEIS A003002 b-file used for window
  bounds, the broad-pass result logs, the recap residuals, the
  HiGHS attack logs, the T1a/T1b/T1c benchmark JSONLs, verified DRAT
  certificates, LRAT artifacts where emitted with \texttt{drat-trim\ -L}, the Lean T1c
  proof-search targets, environment/version captures for the main solver stack,
  and the scripts needed to generate T3 are all included.
\item
  Existing single-instance entry points (\texttt{r3\_split\_cpsat.py} and
  \texttt{r3\_highs\_attack.py}, and \texttt{r3\_sat\_attack.py}) can rerun any row of T1c,
  T1b, or T2 by chunk ID.
\end{itemize}

The large solver artifacts are archived on Zenodo at
\url{https://doi.org/10.5281/zenodo.20463334}. This archive contains the
CNF/DRAT proof artifacts for the \texttt{18} CDCL-closed T1b ~T1c chunks,
the available LRAT outputs, verification summaries, SLURM logs, solver
outputs, SHA256 manifests, and reconstruction instructions for split
archives \cite{ergezer-r3-212-artifacts-2026}.

We welcome correspondence on partial results: any solver run that
closes a strict subset of T1 or T2 is a meaningful incremental
contribution, even if the full upper bound remains open.

\section{6. Discussion}\label{discussion}

We close with three observations: what the architecture transfers to,
where it breaks down, and what the campaign suggests about the
broader proof-search problem class for \texttt{r\_3} upper bounds.

\subsection{\texorpdfstring{6.1 Reusability for adjacent \texttt{N}}{6.1 Reusability for adjacent N}}\label{reusability-for-adjacent-n}

The five-component architecture of §2 is parametric in \texttt{N} and \texttt{K}
and depends on two external inputs: a verified \texttt{(K\ -\ 1)}-element
lower-bound witness for \texttt{r\_3(N)} and the OEIS A003002 b-file prefix
for window-cardinality pruning at lengths \texttt{L\ ≤\ N\ -\ 1}. Both inputs
are standard finite data at the current frontier: exact values through
\texttt{N\ =\ 211} are tabulated by Cariboni/OEIS A003002, and any new run also
requires an explicit verified lower-bound witness at the target size.
We expect the architecture to apply directly to
\texttt{N\ ∈\ \textbackslash{}\{213,\ …,\ 220\textbackslash{}\}} once such witnesses are supplied, with three caveats:

\begin{enumerate}
\def\labelenumi{\arabic{enumi}.}
\tightlist
\item
  The depth of the broad split (\texttt{d\ =\ 24} for \texttt{N\ =\ 212}) is empirical
  and likely needs to grow with \texttt{N}. The number of feasible-after-pruning
  chunks at depth \texttt{d} scales roughly geometrically with the size of the
  witness, so the broad layer for larger \texttt{N} will emit more chunks, and
  per-chunk cost will also rise as the number of variables and 3-AP triples
  grows.
\item
  The \texttt{r\_3(N)} value drops slowly as \texttt{N} grows, so the equality
  \texttt{sum\ x\_i\ =\ K} becomes harder to refute by simple cardinality
  arguments. We expect the \texttt{UNK} rate at the \texttt{60}-s wall cap to
  grow with \texttt{N}.
\item
  The endpoint-forcing argument generalizes: if \texttt{r\_3(N\ -\ 1)\ =\ K\ -\ 1},
  then any \texttt{K}-element 3-AP-free subset of \texttt{{[}1,\ N{]}} contains both
  endpoints. This is the only structural input from prior work that
  the architecture exploits; everything else is \texttt{N}-uniform.
\end{enumerate}

Within the OEIS A003002 frontier, the architecture is therefore a
drop-in tool for any incremental \texttt{r\_3(N\textquotesingle{})} upper-bound attempt, with
the same caveat the present campaign documents: the hard pocket of §4
will likely have an analogue at larger \texttt{N}, possibly more severe.

\subsection{6.2 Limitations}\label{limitations}

The campaign's main limitation is the unresolved hard pocket itself.
Even granting all five CP-SAT levers and the HiGHS substitution, the
architecture cannot close \texttt{r\_3(212)\ ≤\ 43} without an additional idea.
Section 5 frames this as an open problem; here we record three
narrower limitations of the present implementation:

\begin{itemize}
\tightlist
\item
  \textbf{Machine-checkable certificates cover the CDCL-resolved T1b ∖ T1c chunks,
  but not the full campaign.} Follow-up proof-producing runs emitted DRAT
  proofs for all \texttt{18} CDCL-resolved T1b ∖ T1c chunks, and independent
  \texttt{drat-trim} verification confirmed all \texttt{18\ /\ 18}. The \texttt{25} HiGHS-closed T1a
  chunks and the \texttt{\textasciitilde{}93,929} broad-pass CP-SAT INFEASIBLE returns remain
  solver-attested rather than third-party-verified; closing this remaining
  gap would require either a SAT re-encoding of each chunk or a verified
  LP-duality-style certificate format we are not aware of in the open MIP
  ecosystem.
\item
  \textbf{Compute coverage is incomplete.} The campaign processed
  \texttt{100,000} of the \texttt{12,582,912} AP-pruned depth-\texttt{24} chunks at the
  broad layer, plus refinements on small samples. A full sweep would require
  orders of magnitude more worker-hours under the present parameters and was
  not justified by the available evidence, given the hard-pocket diagnosis.
\item
  \textbf{Witness dependence is structural, not adversarial.} The broad
  split is anchored to one specific \texttt{43}-witness. A different
  witness would induce a different chunk-ID space and possibly a
  different hard pocket. We did not test whether the hard pocket
  is witness-invariant; if it is not, alternative witnesses might
  bypass the T1c residual of §4.3.
\item
  \textbf{The Lean benchmark is not yet upstreamed.} We did not find a public
  A003002 / \texttt{r\_3(N)} entry in \texttt{google-deepmind/formal-conjectures} as of the
  campaign date. The repository includes a Lean 4 formalization of the witness
  and T1c targets, but upstreaming and independent typechecking in that
  ecosystem remain follow-on work.
\end{itemize}

The third limitation is the most interesting follow-on: a witness-
ensemble version of the broad split, in which the chunk space is
the disjoint union of split-prefixes from multiple distinct
\texttt{43}-witnesses, would either reveal an easier alternative covering
of the search space or confirm that the hard pocket is intrinsic
to the constraint family rather than to one witness.

\subsection{\texorpdfstring{6.3 What the campaign suggests about \texttt{r\_3} upper-bound search}{6.3 What the campaign suggests about r\_3 upper-bound search}}\label{what-the-campaign-suggests-about-r_3-upper-bound-search}

The dominant finding is now more precise than the original CP-SAT/HiGHS
comparison. LP-paradigm methods --- CP-SAT-style propagation on the bounded
model and HiGHS LP-MIP --- fail uniformly on the \texttt{20} T1b chunks with flat dual
bounds, while CDCL closes \texttt{18\ /\ 20} of them. The residual T1c set has size
\texttt{2} and is the only audited pocket that remains resistant across every tested
paradigm.

Two narrower suggestions follow from the lever inventory of §4.4.
First, the window-cardinality family from OEIS A003002 is doing
nearly all of the work that generic constraint propagation can do;
the \texttt{\textasciitilde{}28}-percentage-point reduction it produces is not matched by
any other intervention at any cost. Second, the levers that \emph{should}
have moved the residual under standard MIP/CP intuition --- variable
reordering, pair-AND Tseitin propagators, walltime extension to the
plateau --- moved it by less than two percentage points each. The
residual is not lying in wait for a smarter generic search.

The exception is the CDCL paradigm switch (§4.3), which closes \texttt{18\ /\ 20} of
T1b. We treat that as a qualitative change of proof system, not another lever
inside the LP-relaxation family.

Combined, these suggest that the most productive next experiment is
not another CP-SAT or MIP variant but either (i) a stronger SAT/proof-search
attack on T1c, (ii) a tighter upper-bound family on \texttt{sum\ x\_i}
(Fourier-analytic, multi-window partition, or SDP-derived; §5.4), or
(iii) a custom branch-and-bound that hard-codes the reflection symmetry and
any per-instance dominance relations the generic solvers cannot infer. The
benchmark release of §5 is built to support exactly this kind of follow-on.

We end with the working conjecture of §4.5: T1c corresponds to a
low-dimensional pocket in the depth-\texttt{24} assignment lattice along which the LP
relaxation gap is tight and the integer infeasibility certificate, while
present, is not learned by current CDCL encodings under the tested wall caps. A
proof or refutation of this conjecture would be the cleanest mathematical
outcome of the campaign that does not require closing the full unit gap
directly.

\section*{Acknowledgments}
This work was performed on the Unity High Performance Computing (HPC)
platform, a collaborative, multi-institutional resource supported by the
Massachusetts Green High Performance Computing Center (MGHPCC) and its member
institutions.  We acknowledge OEIS contributor Cariboni for the A003002 b-file
at the current frontier $N \leq 211$.

\bibliographystyle{amsalpha}
\bibliography{refs}

\end{document}